\documentclass[namedreferences]{SolarPhysics}
\usepackage[optionalrh]{spr-sola-addons} 
\usepackage{graphicx} 
\usepackage{url}                         

\newcommand{\citet}[1]{\citeauthor{#1} (\citeyear{#1})}
\newcommand{\citep}[1]{\citeauthor{#1} \citeyear{#1}}

\newcommand{\aap}{    {\it Astron. Astrophys.}}
\newcommand{\aaps}{   {\it Astron. Astrophys. Suppl.}}

\newcommand{\aj}{     {\it Astron. J.}} 
\newcommand{\apj}{    {\it Astrophys. J.}}
\newcommand{\apjl}{   {\it Astrophys. J. Lett.}}
\newcommand{\apjs}{   {\it Astrophys. J. Suppl.}}
 
\newcommand{\cjaa}{   {\it Chin. J. Astron. Astrophys.}} 

\newcommand{\grl}{    {\it Geophys. Res. Lett.}}

\newcommand{\mnras}{  {\it Mon. Not. Roy. Astron. Soc.}}

\newcommand{\pasj}{   {\it Publ. Astron. Soc. Japan}}

\newcommand{\solphys}{{\it Solar Phys.}}
 
\newcommand{\ssr}{    {\it Space Sci. Rev.}}

\begin{document}

\begin{article}

\begin{opening}

\title{Magnetic Field Extrapolations into the Corona: Success and Future
Improvements}

%
\author{S.~\surname{R\'egnier}$^{1}$}

%
\runningauthor{S. R{\'e}gnier}
\runningtitle{Extrapolation in the Corona}

%
  \institute{$^{1}$ Jeremiah Horrocks Institute, University of Central
  Lancashire, Preston, Lancashire, PR1 2HE, United Kingdom
                     email: \url{SRegnier@uclan.ac.uk} 
             }

\begin{abstract}
The solar atmosphere being magnetic in nature, the understanding of the
structure and evolution of the magnetic field in different regions of the
solar atmosphere has been an important task over the past decades. This task
has been made complicated by the difficulties to measure the magnetic field
in the corona, while it is currently known with a good accuracy in the
photosphere and/or chromosphere. Thus, to determine the coronal magnetic field, a
mathematical method has been developed based on the observed magnetic
field. This is the so-called magnetic field extrapolation technique. This
technique relies on two crucial points: (i) the physical assumption leading to
the system of differential equations to be solved, (ii) the choice and quality
of the associated boundary conditions. In this review, I summarise the physical
assumptions currently in use and the findings at different scales in the solar
atmosphere. I concentrate the discussion on the extrapolation techniques
applied to solar magnetic data and the comparison with observations in a broad
range of wavelengths (from hard X-rays to radio emission).

\end{abstract}

%

\end{opening}

%

\section{Introduction}\label{sec:intro}

With the advent of photospheric magnetographs/magnetometers in the
sixties, it has soon been realised that the coronal magnetic field can be
derived by assuming an equilibrium state. The main forces excerted on the coronal
plasma are the
plasma pressure, the gravitational force, and the magnetic or Lorentz forces.
This gives the magnetohydrostatic equilibrium:
\begin{equation}
-{\bf \nabla}p + \rho {\bf g} + {\bf j} \times {\bf B} = {\bf 0}
\end{equation} 
where $p$ and $\rho$ are the plasma pressure and density, ${\bf g}$ is the
gravity, ${\bf j}$ is the electric current density, and ${\bf B}$ is the magnetic
field. This equation is a partial different equation, which can be solved by
imposing a set of boundary conditions in a finite or semi-finite domain of
computation. In addition, Maxwell's equation given the divergence-freeness
of the magnetic field has also to be considered. The technique is the so-called
magnetic field extrapolation/reconstruction (nowadays both words are used
without distinction). Two problems have to be distinguished: (i) to numerically
solve the system of equations, (ii) to incorporate solar data into the numerical
codes. Both problems need a careful treatment as the convergence of a
well-established numerical code does not imply the convergence of the same code
with a specific dataset. In this review, I will address the magnetic field
extrapolations performed using photospheric/chromospheric magnetograms as
boundary conditions. This indeed assumes that the algorithms to extrapolate the
magnetic field are behaving well with data.

The magnetograms used as boundary conditions are recorded by ground-based
or space-born instruments. For this review, the pros and cons of different
instruments are not taken into account and only the physics problems and
results addressed in the cited papers are considered. The aim of this review
is to show that the magnetic field extrapolation techniques have been used to
tackle successfully a broad range of solar physics issues.  

Recently, two reviews have been published with a different focus to this paper:
(i) \citet{wie12} focused their discussion on the nonlinear force-free field
methods with a short section on the comparison with the structures of the
solar corona, (ii) \citet{lrsp-2012-6} describe the progress in the modelling of
global photospheric and coronal magnetic fields which includes extrapolation
techniques as well as flux transport models which are not
discussed here.

The review is organised as follows. I first give a brief review of the different
models used to reconstruct the solar atmospheric magnetic field (see
Section~\ref{sec:model}). In Section~\ref{sec:phys}, the important quantities
that can be derived from a magnetic field configuration are listed. Thus, the
results obtained from magnetic field extrapolations are discussed for the quiet
Sun (see Section~\ref{sec:qs}), and for active regions (see
Section~\ref{sec:ar}). To conclude (see Section~\ref{sec:disc}), I discuss the
issues which are still to be resolved, and the futher developments that I
envision for the magnetic field extrapolation techniques.

\section{A Brief Review of Numerical Models} \label{sec:model}

Magnetic field extrapolation techniques aim at solving a system of differential
equations with the appropriate set of boundary conditions. An extrapolation scheme is
judged on the well-posedness of the problem, that is to say to find the right
combination of boundary conditions to solve a particular physical problem. I
will thus distinguish between (i) the goodness of the numerical scheme which is
usually tested with analytical and/or semi-analytical solutions, and (ii) the
goodness of the extrapolation which includes the effects of the boundary
conditions as provided by observations. Most of the schemes described below are
good numerical schemes; however, their behaviour when extrapolating solar data
can be significantly different, or has not yet been tested. 

	\subsection{Potential Fields}

The potential field is the solution of Laplace's equation:
\begin{equation}
{\bf \nabla}^2 {\bf B} = 0. 
\end{equation}
The solutions of this equation are well-known as harmonic functions in different
geometries. The potential field is relatively easy to compute as it only
requires the knowledge of the normal component of the magnetic field on the
boundaries of the domain. For instance, to compute the potential field in a
Cartesian box only the normal/vertical component is needed on the bottom
boundary as often provided by line-of-sight magnetic field measurements, whilst
closed/open boundary can be imposed on the side and top boundaries. The first
potential field extrapolation techniques have been in the early 60s with the
development of photospheric magnetographs 
\cite{1964NASSP..50..107S,1969SoPh....9..131A,1974SoPh...36..345L,%
	1969SoPh....6..442S,1976SoPh...46..185A,1982SoPh...77..363L,%
	1995SoPh..159...89H,2001SoPh..198....5R}. Several methods have been
compared by \citet{1982SoPh...81...69S} showing that the connectivity
and geometry of field lines can be different fom one model to the other.
 
	\subsection{Force-free Fields}

We usually distinguish between potential field and force-free field
extrapolations even if the potential field is a particular case of force-free
field, and thus has similar properties of existence and stability \cite{mol74}. 
The force-free fields assume implicitely that the solar atmospheric plasma is a
low $\beta$ plasma.

		\subsubsection{Linear Force-free Field}

The linear force-free (LFF) field is described by the following equation:
\begin{equation}
{\bf \nabla} \times {\bf B} = \alpha {\bf B}
\end{equation}
where $\alpha$ is a scalar called the force-free parameter. The boundary
conditions are given by the normal component of the magnetic field on the
boundaries of the computational box, to which a guess for the $\alpha$-value is
added. The numerical methods to solve this linear problem include vertical
integration, Green's functions, Fourier transforms, spherical harmonics, and
boundary integrals in either Cartesian or spherical coordinates	
(\opencite{1972SoPh...25..127N};
\opencite{1974SoPh...36..345L};
\opencite{1975AN....296..177S};
\opencite{1977ApJ...212..873C};
\opencite{1978A&A....69...43N};
\opencite{1981A&A...100..197A};
\opencite{1984AcApS...4..272W}, \citeyear{1985ChA&A...9...44W};
\opencite{1988A&A...198..293S};
\opencite{1989AuJPh..42..317D};
\opencite{1989ApJS...69..323G};
\opencite{1995SoPh..159...97Y};
\opencite{1996SoPh..168...47A};
\opencite{1996ApJ...461..415K};
\opencite{1998A&A...339..252A};
\opencite{1999SSRv...87..145C};
\opencite{2000A&A...361..743C}).
Very few studies of the behaviour of LFF field algorithms have
been performed \cite{2009SoPh..260..109L}.

As it is obvious from photospheric vector magnetic field observations that a
single value of $\alpha$ cannot describe the coronal magnetic field 
\cite{1999SoPh..188....3L,1999SoPh..188...21L}, several
methods have been attempted to derive the geometry of a single coronal loop
using the $\alpha$-value that will best fit the observed loop
\cite{2002SoPh..208..233W,%
2003SoPh..218...29C}. In other words, this
corresponds to a piecewise LFF field for a discrete (usually small) number of
loops.   	

		\subsubsection{Nonlinear Force-free Field}

The nonlinear force-free (NLFF) field satisfies the following equation:
\begin{equation}
{\bf \nabla} \times {\bf B} = \alpha({\bf r}) {\bf B},
\end{equation}
where $\alpha$ is the force-free function depending on the position ${\bf r}$.
Taking the divergence of the above equation, we obtain that
\begin{equation}
{\bf B} \cdot {\bf \nabla} \alpha = 0
\end{equation}
implying that $\alpha$ is constant along a given magnetic field line. This fact
is reinforced in the corona as the conduction along the magnetic field dominates
(the conduction accross field lines is negligible). The NLFF methods have been
summarised by \citet{wie12}. The main methods to extrapolate the magnetic field
into the corona as a NLFF field are:
\begin{itemize}
\item[- Vertical integration:]{the method consists in propagating the boundary
conditions into the corona from the bottom boundary, layer by layer
(\opencite{1985ChA&A...9...44W};
\opencite{1989A&A...216..265C}, \citeyear{1990A&A...227..583C}, 
\citeyear{1990A&A...230..193C};
\opencite{1990ApJ...362..698W};
\opencite{1992A&A...263..351D};
\opencite{2006ApJ...649.1084S}, 
\citeyear{2007ApJ...666..491S}).}
\item[- MHD evolutionary techniques:] {based on the low plasma-$\beta$ MHD
equations, an initial configuration including electric currents is relaxed to a
nonlinear force-free state owing to resistivity, also know as stress-and-relax
models \cite{1986ApJ...309..383Y,%
	1988ApJ...328..830M,%
	1990CoPhC..59...21S,%
	1994ApJ...422..899M,%
	1996ApJ...473.1095R,%
	2005A&A...433..335V,2007SoPh..245..263V,2010A&A...519A..44V,%
	2011ApJ...727..101J}.}
\item[- Optimization:]{the basic principle is to minimise a functional containing the
force-free constraint as well as the solenoidal constraint to relax an initial
configuration towards a nonlinear force-free state
(\opencite{2000ApJ...540.1150W};
\opencite{2004SoPh..219...87W};
\opencite{2005SoPh..228...67W}, \citeyear{2008SoPh..247..249W},
\citeyear{2010A&A...511A...4W};
\opencite{2006SoPh..233..215W};
\opencite{2006SoPh..235..201I};
\opencite{2006A&A...457.1053W};
\opencite{2009A&A...508..421T};
\opencite{2009Ge&Ae..49..940M};
\opencite{2010A&A...516A.107W};
\opencite{2011A&A...526A..70F}). In the recent years, the optimization scheme has
	been developed to include more constraints and thus to obtain an
	equilibrium closer to a force-free field.}
\item[- Grad--Rubin methods:] {the nonlinear force-free equation being a system of
partial differential equations of mixed type, the method described by
\citeauthor{gra58} \shortcite{gra58} consists in separating the elliptic part
(force-free equation) and the hyperbolic part (gradient of $\alpha$) of the
system, each system being then linear and easier to solve 
\cite{gra58,1981SoPh...69..343S,1997SoPh..174..129A,%
	1999A&A...350.1051A,2004SoPh..222..247W,%
	2006A&A...446..691A,2006SoPh..235..201I,2006SoPh..238...29W,%
	2009ApJ...700L..88W,2009ApJ...707.1044M,2010A&A...522A..52A,2012ApJ...756..153M}.
	 The Grad--Rubin method requires as boundary
	conditions the vertical/radial component of the magnetic field on each
	boundary, as well as the distribution of $\alpha$ in a chosen
	polarity. This ensures that the extrapolation method is a mathematically
	well-posed problem in Hadamard sense.}
\item[- Boundary integral:] {the nonlinear force-free model can be describe as
an exterior problem and a boundary integral equation written in which the
magnetic field component within a volume can be determined by the magnetic field
components on the boundary. The formulation of the boundary integrals can be
found by \citet{1963JAM....30..158C} and recently applied to solar cases
(\opencite{1995AcApS..15..359W};
\opencite{1997SoPh..174...65Y}, \citeyear{2000SoPh..195...89Y};
\opencite{2004MNRAS.347.1255L};
\opencite{2006MNRAS.369..207H};
\opencite{2006ApJ...638.1162Y};
\opencite{2008JGRA..11305S90H}).}
\item[- Force-free electrodynamics:]{the theory of force-free electrodynamics is
applied to the modelling of coronal magnetic fields which has been applied
successfully to pulsar magnetospheres \cite{2011SoPh..269..351C,%
	2013SoPh..282..419C}.}
\end{itemize} 

With the development of different numerical techniques, testing and
comparing different algorithms has been proven to be needed. The main
efforts have been done since 2004 by the international group on nonlinear
force-free modelling (\opencite{2006SoPh..235..161S}, 
\citeyear{2008ApJ...675.1637S};
\opencite{2008SoPh..247..269M};
\opencite{2009ApJ...696.1780D}). One aim of this series of papers
has been to show that modellers need to be careful when injecting the boundary
conditions into the numerical codes: the boundary conditions have to be
consistent with the assumption of the model and the set-up of the numerical
code. For instance, in \inlinecite{2008ApJ...675.1637S}, it has been shown that
by selecting carefully the photospheric magnetic field, the coronal magnetic
field structure in the core of the reconstructed active region was similar for
all NLFF methods. Several other papers have addressed quantitative differences
between several algorithms 
\cite{2006SoPh..235..201I,2006ApJ...641.1188B,2009SoPh..257..287R,2009Ge&Ae..49..940M,%
2011SoPh..269...41L}.  
	
		\subsection{Non-force-free Fields}

The next step after the force-free field extrapolation is to solve the
magnetohydrostatic (MHS) equilibrium equation, which includes the plasma pressure
force and the gravitational force in addition to the magnetic forces. An
algorithm has been developed by \citet{gra58} as a well-posed problem. This
algorithm has been recently implemented by \citet{2013SoPh..282..283G}
neglecting the gravity (see also \citeauthor{2009arXiv0911.5214B}
\citeyear{2009arXiv0911.5214B}). The optimization scheme has also been adapted
to solve the magnetohydrostatic equation
\cite{2006A&A...457.1053W,2007A&A...475..701W}. The models usually require to
define a realistic atmosphere from the photosphere to the corona. As a first order
assumption (neglecting the feedback between the magnetic field and plasma
properties), \citet{reg08a} derived the global properties of Alfv\'en speed,
plasma $\beta$-values and reconnection rate in the corona by assuming a nonlinear
force-free equilibrium for the magnetic field and a hydrostatic equilibrium for
the plasma. 

A minimum dissipation rate method has been adapted to include photospheric
measurements (\opencite{2006GeoRL..3315106H}, \citeyear{2008SoPh..247...87H};
\opencite{2007SoPh..240...63B};
\opencite{2008ApJ...679..848H},
\citeyear{2010JASTP..72..219H}). The method
describes a general or non-force-free magnetic field by a superimposition of one
potential field and two linear force-free fields.

\citet{1999SoPh..186..123G} have developed a non-force-free model, the so-called
stretched magnetic field, which combined photospheric magnetograms and the
structures of coronal loops.  

		\subsection{Miscellaneous}

I will also mention several other analytical/numerical methods that have been
used to access the structure of the three-dimensional coronal magnetic field,
but which cannot be classified as extrapolation techniques as such.

This first method to have produced significant results is the point-charge
method. A magnetogram is decomposed in a discrete number of polarities, which
are
reduced to points associated with the magnetic flux of the polarity (zero
magnetic field outside the point charges). The magnetic field is thus given by
an analytical solution. For instance, the method has been developed and used for
theorising  the topology of magnetic field in different geometries
(\opencite{1993A&A...276..564T};
\opencite{1996A&A...308..233B};
\opencite{1999RSPSA.455.3931B};
\opencite{2002ApJ...579..468L};
\opencite{2002SoPh..209..333B}, \citeyear{2004GApFD..98..429B};
\opencite{2005ApJ...629..561B};
\opencite{2006SoPh..237..227M};
\opencite{2006SoPh..238...13M}).
The point-charge method has also been used to fit the observed 3D coronal loops
following a nonlinear force-free assumption (\opencite{2012SoPh..tmp..181A}, 
\citeyear{2012SoPh..tmp..329A}, 
\citeyear{2013ApJ...763..115A};
\opencite{2012SoPh..tmp..182A}).
Another successful model to determine the properties of the coronal magnetic
field is the flux rope insertion model developed by \citet{2004ApJ...612..519V}.
Starting from an equilibrium state (often a potential field), a flux rope or
twisted flux tube is inserted within the magnetic configuration by modifying the
boundary conditions and by matching the observed X-ray sigmoid or H$\alpha$
filament (see {\em e.g.}, \opencite{2007JASTP..69...24V}).

		\subsection{Boundary Conditions}
		
In order to solve a system of differential equations, it is important to define
the correct boundary conditions that will lead to a mathematically well-posed
problem (or not) in the Hadamard sense. 

For the potential field, only the normal component of the magnetic field
on the boundaries of the computational box is needed. 

For the linear force-free field, the normal component is required, to which a
guess for the value of $\alpha$ in the computational volume will be added.

For the nonlinear force-free field, the normal component of the magnetic field
as well as the distribution of $\alpha$ has to be imposed. According to
\citet{gra58}, imposing the distribution of $\alpha$ in one chosen polarity as
the boundary condition will lead to a mathematically well-posed problem. It is
important to notice here that in order to compute the distribution of $\alpha$
from vector magnetograms, the following formula is used (in Cartesian
coordinates with $(x, y)$ on the photospheric surface):
\begin{equation}
\alpha = \frac{1}{B_z}\left(\frac{\partial B_y}{\partial x} - \frac{\partial
B_x}{\partial y} \right). \label{eq:alpha}
\end{equation}  
This equation is considered owing to the measurement of the magnetic field vector
on the $xy$-plane, implying that no derivatives with respect to $z$ can be
computed. A further discussion on the implication of this formula will be done
in Section~\ref{sec:Cbound}.

For the minimum dissipation model, the vector magnetic field components on one or two
different atmospheric layers are needed. 

For the magnetostatic model, the same boundary conditions as for the non-linear
force-free model are used in addition to prescribing the plasma pressure on the
photospheric level in one chosen polarity. If the gravitational force is
considered, then an initial atmosphere model will be needed.

\section{Physical Quantities} \label{sec:phys}

Below, I list the important quantities that can be derived from the
magnetic field extrapolations and which will lead to an in-depth physical
interpretation of the magnetic field structure and coronal phenomena as
described in Sections~\ref{sec:qs} and~\ref{sec:ar}.

	\subsection{Magnetic Energy} \label{sec:energy}
	
\paragraph*{\em Magnetic Energy} The magnetic energy computed in a volume $V$
is given by
\begin{equation} \label{eq:em}
E_{\textrm{m}} = \int_V\, \frac{B^2}{2\mu_0}~dV.
\end{equation}
The reconstruction models provide the three components of the magnetic field at
discrete locations in the computational volume. The magnetic energy can then be
easily computed by discretizing Equation~(\ref{eq:em}). The magnetic energy for the
potential field is a minimum (lower bound) of magnetic energy for a force-free field
computed with the same boundary conditions, {\em i.e.}, the same vertical/normal magnetic
component on all boundaries. The LFF magnetic energy is a minimum of
magnetic energy for the NLFF field if the magnetic helicity is
conserved \cite{wol58,tay74}. 

\paragraph*{\em Free Magnetic Energy} The free magnetic energy is the
difference in magnetic energy for a magnetic field model and for a reference
field:
\begin{equation}
\Delta E_{\textrm{m}} = E_{\textrm{mod}} - E_{\textrm{ref}}.
\end{equation}
The reference field is often chosen to be the potential magnetic field as it is
the minimum energy state \cite{aly84}. The free magnetic energy is a measure of the
magnetic energy that can be stored in the magnetic configuration or released
during an eruptive or reconnection event. 

\paragraph*{\em Magnetic Energy Density} The magnetic energy density is the
term $\frac{B^2}{2\mu_0}$ taken at one particular location within the volume $V$.
The energy density can also be seen as the magnetic energy in a small volume
$\delta V$ ({\em e.g.}, a single pixel). As such, the magnetic energy density
provides the distribution of magnetic energy in a small volume. However, the free
magnetic energy density is of no use/meaning as the connectivity of magnetic
field lines between the magnetic field and a reference field can be drastically
different.  

\paragraph*{\em The Aly-Sturrock Conjecture} Both \citeauthor{aly89}
(\citeyear{aly89}) and \citeauthor{stu91} (\citeyear{stu91}) showed that there
exists an upper bound of the magnetic energy of force-free fields when the field
is totally open (unipolar field). The condition of application of this
conjecture is when the magnetic field within the volume decays rapidly towards
the boundaries, or when the magnetic flux through the boundaries other than the
bottom boundary becomes negligible. This condition is easily satisfied when considering the half
space above the photosphere, or when the boundaries are far
enough from the magnetic field sources. This condition is not satisfied when
magnetic flux is present near the edges of the boundaries.

\paragraph*{\em Time Evolution} One of the most interesting physical issues
about magnetic extrapolation has been to know if the reconstructed magnetic
field can describe the time evolution of solar regions. In the non-eruptive
solar corona, the time of evolution is mostly given by the Alfv\'en transit
time along individual magnetic field lines or coronal loops for active regions,
and by the comparison of Alfv\'en transit time and granular motion for the quiet
Sun. For active regions, the Alfv\'en transit time of a typical loop of length
200 Mm is about 10-15 min; therefore if there is no major injection of magnetic
energy or magnetic helicity, the evolution of active regions can be studied by a
time series of equilibria.
	
	\subsection{Magnetic Helicity}

\paragraph*{\em General Definition} The concept of magnetic helicity has been
largely used in plasma physics to describe the complexity of the magnetic field,
including the twist and shear in magnetic field lines. Maxwell's equation, 
\begin{equation}
{\bf \nabla} \cdot {\bf B} = 0
\end{equation}
implies that the magnetic field is solenoidal and can be
described by a vector potential ${\bf A}$ such that ${\bf B} = {\bf \nabla} \times
{\bf A}$. The definition of ${\bf A}$ is not
unique and depends on a gauge condition. From this definition, the magnetic
helicity in a volume $V$ is defined as
\begin{equation}
\label{eq:hel_mag}
H_{\textrm{m}} = \int_{V}\,{\bf A} \cdot {\bf B}\,dV = \int_{V}\,{\bf A} \cdot  {\bf \nabla}
\times {\bf A}\,dV.
\end{equation}
The main concepts about magnetic helicity have been summarised by
\citet{1999PPCF...41..167B}.

\paragraph*{\em Relative Magnetic Helicity} As the definition of
Equation~(\ref{eq:hel_mag}) is not unique, a gauge invariant definition has been
developed by \citeauthor{ber84} (\citeyear{ber84})
\begin{equation}
\label{eq:rhel_bf}
\Delta H_{\textrm{m}} = \int_{V}\,({\bf A} - {\bf A_{\textrm{ref}}}) \cdot 
({\bf B} + {\bf B_{\textrm{ref}}})\,dV
\end{equation} 
and also by \citeauthor{fin85} (\citeyear{fin85})
\begin{equation}
\label{eq:rhel_fa}
\Delta H_{\textrm{m}} = \int_{V}\,({\bf A} + {\bf A_{\textrm{ref}}}) \cdot 
({\bf B} - {\bf B_{\textrm{ref}}})\,dV.
\end{equation}
The definition of \citeauthor{ber84} (\citeyear{ber84}) originally contained a surface term, which
tends to zero when considering the half space above the photosphere (or
boundaries far away from the strong magnetic field regions) or vanishes for a
finite volume when the following boundary conditions are imposed: the magnetic
field normal to the surface of the finite volume is the same for both magnetic
fields, the divergence of the reference vector potential vanishes, and the
reference vector potential  is perpendicular to the surface of the finite
volume. In those two cases, the \citeauthor{fin85} (\citeyear{fin85}) and 
\citeauthor{ber84} (\citeyear{ber84}) formulae are equivalent.   
 
\paragraph*{\em Magnetic Helicity Conservation} In ideal MHD, the magnetic
helicity is invariant during the evolution of any closed flux system
(\citeauthor{wol58} \citeyear{wol58}). \citet{tay74} applied  this to laboratory
experiments and hypothesized that, for a weak but finite resistivity, the total
magnetic helicity of the flux system is invariant during the relaxation process.
Taylor's theory has often been invoked to assume that the magnetic helicity in a
solar atmospheric region is conserved. This latter statement is valid only if
the solar region under consideration is a closed flux system. For instance, active
regions in which flux emergence is taking place or a CME has originated cannot
satisfy helicity conservation. The redistribution of magnetic helicity at large
scale in the solar atmosphere or in the heliosphere is a basic consequence of
the magnetic cycle and the sustainability of the dynamo action. 

\citeauthor{wol58} (\citeyear{wol58}) has shown that the minimum energy of a
NLFF field is a LFF field when the magnetic helicity is conserved ({\em i.e.}, for a
closed system). This implies that the difference between the magnetic energy of
the NLFF field and LFF field is a better estimate of the free magnetic energy
(see Section~\ref{sec:energy}).

\paragraph*{\em Self and Mutual Helicity} The magnetic helicity of a magnetic
configuration can be decomposed into several components: self and mutual
helicity \cite{ber99}, or twist and writhe \cite{ber06}. As an example, for a
single twisted flux tube embedded a uniform external magnetic field, the self
helicity is assumed to correspond to the twist within the flux tube, while the
mutual helicity corresponds to the crossing between the external field and the
twisted flux tube. The definitions given by \citet{ber99} have been studied in
the frame of force-free extrapolation of active regions by \citet{reg05}: the
self helicity is related to the twist of the flux bundles within the active
region, while the mutual helicity contains a contribution from the crossing of
field lines and the large-scale twist of the active region (large compared to
the size of the computational box). 
\citet{2008ApJ...674.1130L} have defined a formula for the self helicity which
describes only the twist of the field lines. 

\paragraph*{\em Twist in Force-free Fields} The magnetic helicity is a measure of
the topology of a magnetic field configuration, and also includes a measure of
the shear and twist of magnetic field lines. In the solar physics literature,
the force-free parameter/function, $\alpha$, is often called the twist. This is
a misnomer as $\alpha$ is strictly equivalent to the twist in a thin flux
tube approximation, which does not always apply to the magnetic field
extrapolated in solar regions. Especially in a non-idealised configuration, the
twist and the $\alpha$-value can have opposite sign 
\cite{reg04,2012SoPh..278..323P}.

\paragraph*{\em Time Evolution} As for the magnetic energy, the evolution of
magnetic helicity injection and redistribution can be studied by a time series
of equilibria. The time scale of the change in magnetic helicity is shorter than
the time scale for the magnetic energy. A general equation for the rate of
change of magnetic helicity has been derived by \citet{1984A&A...137...63H} for a
non-ideal plasma. 

\paragraph*{\em Other Helicities} Magnetic helicity is the integral over a
given volume of the vector potential ${\bf A}$ and its curl, ${\bf B}$. This
definition can be extended to other quantities such as the electric current
density or the vorticity. The current helicity is 
\begin{equation}
H_{\textrm{c}} = \int_V\, {\bf j} \cdot {\bf B}\,dV
\end{equation}
with ${\bf \nabla} \times {\bf B} = \mu_0  {\bf j}$, and the hydrodynamical helicity
by
\begin{equation}
H = \int_V\,{\bf v} \cdot {\bf \omega} \,dV
\end{equation}
where ${\bf v}$ is the flow field and ${\bf \omega} = {\bf \nabla} \times  {\bf v}$
is the associated vorticity.

From vector magnetograms, the current helicity on the surface has been estimated
either by assuming that the transverse components of the magnetic field and/or
the electric current density are negligible, or by assuming a LFF field ({\em 
e.g.},
\citep{1996SoPh..168...75A}). For the
latter assumption, the current helicity density is $h_{\textrm{c}} = \alpha B^2$.
For a volume $V$, the current helicity associated with a LFF field is 
\begin{equation}
H_{\textrm{c}} = \int_V\,\alpha B^2\,dV = 2 \mu_0\,\alpha\,E_m
\end{equation}
where $E_{\textrm{m}}$ is the magnetic energy.

	\subsection{Magnetic Topology}
 
As it is beyond the scope of this review to give a complete description of all
topological studies, I will just mentioned that the concept and development of
magnetic topology applied to coronal structures have been reviewed in
\citet{2005LRSP....2....7L}. The transfer of magnetic energy and magnetic
helicity through topological elements is the consequence of magnetic
reconnection, and thus eruptive events. In particular, the main ingredients are
null points where the magnetic field vanishes, separators, separatrices,
quasi-separatrices, and hyperbolic flux tubes.

The magnetic topology of the potential field is assumed to be just slightly
modified compared to other magnetic field models \cite{1999SoPh..186..301H,%
2000SoPh..194..197B}. It has recently been shown by \citet{reg12} that the
location and properties of null points existing in a simple configuration for
potential, LFF, and NLFF fields are similar when the spectral radius (maximum in
the absolute value of the eigenvalues associated with the null point) is large,
meaning that the magnetic null point is embedded in a strong field region or
surrounded by strong electric currents ({\em i.e.}, large magnetic field
gradients).

Null points have been extensively used as a proxy to the complexity of the
magnetic field and the possible existence of reconnection events in a diffusion
region encompassing the null points. This can be extended to the study of
current sheets. The diffusion regions including a null point or degenerated from
a null point are supposed to be the more efficient in releasing magnetic energy
and to convert this energy into kinetic and thermal energies.
\citet{2005ApJ...624.1057P} have pointed out that the magnetic energy storage
induced by slow photospheric motions is much more efficient along separators than
at separatrices. 

Another quantity that has been commonly used is the $Q$-factor or squashing
degree \cite{2007ApJ...660..863T}. The $Q$-factor is a measure of the change of
connectivity between neighbouring field lines and, as such, indicates the
location where the magnetic energy could be dissipated or released. Except few
quantitative studies ({\em e.g.}, \citeauthor{2010SoPh..267..107L} 
\citeyear{2010SoPh..267..107L}), only qualitative
comparisons with observations have been performed.  

\section{Quiet Sun} \label{sec:qs}

\subsection{Validity of the Extrapolation Methods}

The complex physical behaviour of the photosphere and chromosphere makes the
existence of electric currents perpendicular to the magnetic field possible,
especially Hall currents described by the Hall parameter (ratio of the  electron
gyrofrequency to the electron-ion/neutral collision frequency), and the Pedersen
currents (implied by convective electric field). These perpendicular currents
will dissipate rapidly with altitude in the solar atmosphere
\cite{1960MNRAS.120...89G,2000ApJ...533..501G}, leading to a force-free corona
(only with parallel currents). The existence of such electric currents are
crucial for the validity of the extrapolation methods in the quiet Sun. It is
less important for active regions owing to the characteristic time scale of
evolution. 

Extrapolation methods such as potential fields have been applied to the quiet
Sun without a real physical justification, but mostly due to the lack of
observations/boundary conditions which will allow a better physical
description of the system. The quiet-Sun magnetic field extrapolations can be
regarded as a preliminary step towards the small scales of the solar atmosphere.
It is also worth mentioning that the measurement of the magnetic field vector in
the quiet-Sun regions is still challenging.

	\subsection{The Magnetic Carpet}

The nature and structure of the magnetic field in quiet-Sun regions have been
investigated with the means of magnetic field extrapolation. As currently
line-of-sight magnetic field is the only reliable magnetic field component
measured with high accuracy, the potential field model has been used for the
quiet Sun. Note that the interest about the quiet-Sun evolution has recently
been rejuvenated by nonlinear force-free modelling in a theoretical context
\cite{mey11,mey12}.

In \citeyear{1999SoPh..184..239W}, \citeauthor{1999SoPh..184..239W} have used a
potential field model to show that the orientation of the H$\alpha$ fibrils did
not match the potential field lines, and thus concluding that the quiet-Sun
magnetic field contains a non-potential component  (see also
\opencite{2011ApJ...739...67J}). H$\alpha$
fibril orientation has also been compared with linear force-free models by
\citet{1973SoPh...30..421N} finding that there is a value of $\alpha$ giving a
good match between the magnetic field and the fibril direction; however the
height of the structures was the largest discrepancy between the observations
and the model. This fact has been justified by the existence of perpendicular
electric currents in the photosphere and chromosphere such as Pedersen and Hall
currents that will be dispersed with height to vanish in the corona.  The 3D
modelling of the magnetic carpet has been performed based on the point charge
technique \cite{sch02}: the assumption is justified in the quiet Sun as
small-scale magnetic features can be isolated and tracked in time
(\opencite{def07}; \opencite{lam08}, \citeyear{lam10}). \citeauthor{sch02}
(\citeyear{sch02}) showed that EUV brightenings can be correlated with the
change of connectivity in the small-scale magnetic field. Building on this work,
\citeauthor{reg08} (\citeyear{reg08}) performed a potential field extrapolation
(with continuous magnetic field) using a high-resolution {\em Hinode}/SOT
magnetogram to describe the complexity of the quiet-Sun magnetic field. The
authors showed that the complexity of the quiet Sun is located in the
photosphere and chromosphere at altitude less than 2.5 Mm above the surface. In
this modelling, the magnetic field in the quiet Sun has been decomposed into two
components: (i) closed field lines, which characterise the complexity of the
magnetic field (below several hundreds of kilometer according to
\citeauthor{reg08} \citeyear{reg08}) and are linked to granule's boundary and
part of supergranular field, (ii) open field lines (``open" meaning leaving the
box of computation), which is often considered as the source of the solar wind.
The amount of open flux has been measured at different height showing that a
small amount of the photospheric magnetic flux is open above few megameters. The
imbalance of magnetic flux increasing with latitude (towards the poles), the
open magnetic flux increases also with latitude. \citeauthor{jin11}
(\citeyear{jin11}) showed that the open magnetic field does not always originate
from strong polarities (kG field) in the poles using vector magnetic field
measurements from {\em Hinode}/SOT. It is worth noticing that the quiet-Sun
magnetic field, and in particular the funnel structure, has been theoretically
investigated using potential field extrapolation by \citeauthor{aio06}
(\citeyear{aio06}). \\

	\subsection{Eruptive Events}

There are several eruptive events in the quiet Sun or small-scale events which
require high-resolution field extrapolations: blinkers, jets, spicules, Ellerman
bombs, or mini-CMEs. There is few studies tackling the time evolution of the
events themselves, but mostly analysing the structures associated with these
events.

Bright points (often observed in EUV or X-rays) have been studied in details in
term of their magnetic field structure and time evolution. Based on the point
charge technique, \citet{ale11} have shown that the observed structure of the
magnetic field (or some field lines used as a proxy to the magnetic field) is
close to a potential field state.    

With the development of high-resolution instrumentation, the magnetic field at a
granular scale becomes available, and thus also the study of
photospheric and chromospheric phenomena in the quiet Sun or active regions. Based
on the Flare Genesis Experiment observations \cite{ber01}, \citeauthor{par04}
(\citeyear{par04}) have studied the possibility of undulatory emergence in a
small-scale field (a coherent flux tube emerging at different locations
depending on the convection pattern) based on a LFF field extrapolation. The
authors have noticed that the emergence locations are associated with Ellerman
bombs. \citeauthor{gug10} (\citeyear{gug10}) studied reconnection events in a
small-scale magnetic field within an active region during the emergence of
magnetic flux. Using a LFF field extrapolation, the authors have derived the
complexity associated with these events: the small-scale field exhibits a
complex topology with a fan-spine structure.

From a potential extrapolation, \citeauthor{he10} (\citeyear{he10}) have shown
that a jet occurring in a polar coronal hole is guided by the open magnetic field,
and thus can be a source for the fast solar wind.

\section{Active Regions} \label{sec:ar}

Active regions and their related structures and phenomena have been extensively
studied with the advent of line-of-sight magnetographs and vector magnetographs
measuring strong fields (active region fields) with high accuracy.

	\subsection{Structure of the Magnetic Field}

As magnetic field extrapolations are based on an equilibrium, the first step is
to study the static structure of the magnetic field above an active region. 

General properties as follows have been derived:
\begin{itemize}
\item[-]{The magnetic field decays with height, but not following a bipolar
magnetic field relation \cite{reg08a}. The magnetic field can decay by several
orders of magnitude with height.}
\item[-]{Stating the obvious, individual active regions have different
structures in terms of shear and twist, which depends on their history
(\opencite{reg07a}, \citeyear{reg07b}).}
\item[-]{Active regions have a complex distribution of electric current density
generated by photospheric motions and by emergence from below the photosphere
\cite{1996ApJ...462..547L}.}
\item[-]{In a statistical sense, the magnetic field lines are longer and higher
in altitude when electric currents are present \cite{reg07a}.}
\item[-]{The magnetic energy is stored mostly at the bottom of the corona
(including the photosphere and chromosphere), and can also be stored in twisted
flux bundles in the low corona \cite{reg07a}.}
\item[-]{Active regions can be decomposed into two parts: the core, and the
edge. The core is dominated by strong electric current density and thus has
sheared and twisted magnetic field bundles, while the edge of the active region
is less influenced by electric currents and exhibits more potential field lines
\cite{2009ApJ...696.1780D}.}
\end{itemize}

In terms of magnetic energy, active regions have been founded to have a total
magnetic energy between 10$^{31}$ erg to 10$^{34}$ erg mostly depending on the
total unsigned photospheric flux and the electric currents. The magnetic
helicity has a characteristic value of 10$^{42}$ G$^2$ cm$^4$ \cite{reg06}. 

Lots of studies have involved a comparison of the active region structures
obtained using different assumptions. \citeauthor{1999SoPh..186..301H} 
(\citeyear{1999SoPh..186..301H}) and
\citeauthor{reg12} (\citeyear{reg12}) showed that, for a simulated magnetic
configuration, the magnetic topology does not change much between different
force-free models, even if the geometry and the magnetic energy are significantly
modified. 
	
	\subsection{Filament, Sigmoid, and Twisted Flux Bundles} \label{sec:fil}
	
One of the main success of LFF, NLFF, or MHS extrapolation methods has been to
reconstruct twisted flux tubes that unambiguously exist in the corona.

Filaments/prominences are thought to be a coherent collection of twisted magnetic
field lines forming a twisted flux tube/bundle. The force-free field
extrapolation has been used to show that there indeed are. Using a NLFF model,
\citeauthor{yan01} (\citeyear{yan01}) described an active-region filament as a
magnetic rope or twisted flux bundle with three turns. This is a highly twisted
flux tube, which has never been reproduced in magnetic field extrapolation to
date. In order to store plasma/mass, the most plausible structures are magnetic
dips in which magnetic curvature acts against gravity to sustain the plasma
above the surface. Using a LFF approximation, \citeauthor{aul98a}
(\citeyear{aul98a}) have identified a filament by finding the magnetic dips in
the 3D magnetic field configuration \cite{aul98b,aul99}. Based on the same
model, \citeauthor{dud08} (\citeyear{dud08}) showed the complexity of a filament
as it can be fragmented, and not just a single coherent structure, by parasitic
polarities constituting the legs or barbs of the filament. Using a NLFF method,
\citeauthor{reg04} (\citeyear{reg04}) identified a filament as a dipped twisted
flux bundle with a small number of turn (see also
\opencite{2009ApJ...693L..27C}; \opencite{2010ApJ...715.1566C}). \citeauthor{guo10} (\citeyear{guo10})
showed that part of a filament was identified to a twisted flux bundle, while
the rest of the filament was correlated with magnetic dips along sheared
untwisted field lines.  

Sigmoids are expected to play an important role in solar eruptions \cite{can99}.
They are considered as twisted flux bundles.  \citeauthor{reg02}
(\citeyear{reg02}) reproduced a sigmoid as a highly twisted flux tube without
magnetic dips due to the nonuniform distribution of the twist along the
structure. Other studies have been performed to understand the nature and
evolution of X-ray sigmoid as twisted flux tubes \cite{2012ApJ...744...78S,%
2012ApJ...750...15S,%
2012ApJ...747...65I}.

	\subsection{Comparison with Observations}

The most common way of comparing observations and magnetic field extrapolation
is qualitative: overlaying or putting side-by-side images and magnetic
configurations. Some quantitative methods have been developed and are mentioned
in the following paragraphs.
	
\paragraph*{Optical Wavelengths} The main feature observed in the visible
wavelengths is a
filament mostly in the H$\alpha$ line. The findings using field extrapolations
are summarised in the previous section. Structures like fibrils are not yet
extrapolated due to their small height in the chromosphere; however their
direction has been used to constrain the orientation of the transverse magnetic
field in vector magnetograms \cite{2008SoPh..247..249W}. During flares, the chromospheric
brightenings (localised patches or ribbons) have been identified as the
footpoints of magnetic field lines linked to the flare process such as
reconnection events or post-flare loop growth. \citeauthor{mas09}
(\citeyear{mas09}) used a potential field extrapolation to show the correlation
between magnetic field lines involved in the flare process and chromospheric
ribbons.  

\paragraph*{\em EUV-UV} Using imagers, the EUV loops have been studied mostly at
171 \AA\ and 193/195 \AA\ due to the characteristics of EUV instruments ({\em 
e.g.},
SOHO/EIT, TRACE, STEREO/SECCHI/EUVI, SDO/AIA). A large number
of studies has been performed to show the consistency of magnetic field
extrapolations with observed EUV loops
\cite{1981A&A....98..316P,2005SoPh..228..339A,%
2007SoPh..241..329Z,2008ApJ...677L.141K,2008SoPh..248..359A,%
2008SoPh..248..379I,2008ApJ...676..672W,%
2009SoPh..259....1S,2010ApJ...715...59C,%
2010AJ....140..723A,2012SoPh..280..475S}. The discrepancy between
potential field lines and EUV loops has been used to show the existence of
nonpotentiality in active regions, {\em i.e.}, the existence of shear and twisted
magnetic fields able to store magnetic energy and to trigger eruptions. Based on
observations from (imaging) spectrometers, the extrapolation methods are
compared to flows/Dopplershifts in active regions \cite{2009ApJ...705..926B,%
2012ApJ...752...13B}.

\paragraph*{\em Soft X-rays} Two main solar features observed in soft X-rays have
been matched to magnetic flux bundles or magnetic field lines obtained from
extrapolation: sigmoids (see Section~\ref{sec:fil}) and X-ray bright points.
The latter have been extensively studied since \citet{1974ApJ...189L..93G}.
Adding magnetic extrapolations, the X-ray bright points have been studied in
more detail \cite{1996SoPh..168..115M,2008ApJ...687.1363B}.  

\paragraph*{\em Hard X-ray} The hard X-ray sources are often observed during
flares and reconnection events. The existence of sources at the footpoints of
eruptive loops is well established, as well as the emission at the flare
loop-top assumed to be located below the current sheet where the magnetic
reconnection occur. The magnetic field extrapolations have helped to better
understand the link between the geometry of loops, the topology of the magnetic
field, and the sources of hard X-ray
\cite{1983SoPh...86..323T,1985SoPh...95..311S,%
2003SSRv..107..111Y,2010ApJ...709L.142X,%
2010A&A...515A...1A,2010ApJ...721L.193L,2012ApJ...746...17G}.  

\paragraph*{\em Radio} The radio emission can be observed during eruptive events
such as flares, CMEs or filament eruptions, and it has been often suggested that
the radio emission in the high corona occurs at the interface between closed and
open magnetic field regions 
(\opencite{1999A&A...343..287K};
\opencite{2000SoPh..193..227B};
\opencite{2001A&A...371..333P};
\opencite{2003A&A...410.1001A}, \citeyear{2005A&A...435.1137A},
\citeyear{2011ApJ...730...57A};
\opencite{2006PASJ...58...55G};
\opencite{2006A&A...454..957A};
\opencite{2006SoPh..239..277Y};
\opencite{2007SoPh..240..107H};
\opencite{2007ChJAA...7..265W};
\opencite{2008ApJ...673L.207N};
\opencite{2008A&A...486..589K};
\opencite{2011ApJ...736...64C};
\opencite{2011ApJ...728....1T};
\opencite{2011A&A...526A.137D};
\opencite{2012ApJ...744..167I};
\opencite{2012A&A...544L..17H}). The gyroemission is also used to determine the
strength of the magnetic field in the corona (\opencite{1981A&A....98..316P};
\opencite{1982SoPh...80..233S};
\opencite{1984A&A...134..185H};
\opencite{1997ApJ...488..488B}, \citeyear{2002ApJ...574..453B};
\opencite{1998ApJ...501..853L}, \citeyear{1999ApJ...510..413L};
\opencite{2000A&AS..144..169G};
\opencite{2005SoPh..226..223R};
\opencite{2012SoPh..276...61B}). The comparison of those
measurements and the magnetic field strength derived from extrapolation has yet
not be proven to be satisfactory: the height of the radio sources does not
always correlate. 

\paragraph*{\em Infrared} Despite the development of solar observations in the
infrared wavelengths, there is, to my knowledge, no study comparing the magnetic
field derived from those observations (including the near-IR He{\sc i} triplet and
Stokes parameter measurements in prominences by \citet{2001A&A...375L..39P})
comparing the extrapolated magnetic field.  
	
	\subsection{Time Evolution}

The characteristic Alfv\'en transit time of an active region loop is of 10-15
min. The time evolution of an active region can thus be studied by a series of
equilibria if the photospheric footpoint motions are ideal MHD motions, which do
not add any topological constraints \cite{1987ApJ...312..886A}. The most
complete study of the evolution of an active region has been performed by
\citet{2012ApJ...748...77S} using a high-cadence, high-resolution time series of
vector magnetograms from SDO/HMI. The authors studied the time evolution before
and after a series of flares showing that the flare changes (slightly) the
magnetic energy; the geometry of the magnetic field lines is also modified
making the magnetic field more confined in low corona. For C-class flares,
\citet{reg06} have performed a study of the evolution of the magnetic energy
and magnetic helicity. The authors showed that the start of the significant
changes occur at about 20 min before the peak of the flare.
\citet{2008A&A...484..495T} showed that the magnetic energy build-up in an
active region can be observed several days prior to the eruption.  
	
	\subsection{Physics of Flares and CMEs}

One of the main early study of a flare using an extrapolation method corresponds
to the Bastille Day flare \cite{2000ApJ...540.1126A}. Using a
LFF approximation, the authors showed that the flaring site was
related to the existence of a null point located in the corona. 

Another aspect of the physics of eruptions is the amount of magnetic energy that
can be released. From a single snapshot (one extrapolation at a given time), the
amount of free energy is estimated using the potential field as a minimum (lower
bound) of magnetic energy. The magnetic energy of a LFF field with
the same magnetic helicity can also be considered as a minimum of magnetic
energy. \citet{reg07b} have demonstrated that the amount of free magnetic energy
is larger than the energy of the flare associated with the active regions
studied (only four active regions with very different structure have been
studied). Numerous studies of flares and CMEs have been combined with magnetic
field extrapolations to support the model of eruption 
(\opencite{1964NASSP..50..107S};
\opencite{1973SoPh...32..173Z};
\opencite{1973SoPh...33..187T};
\opencite{1975SoPh...41..397R};
\opencite{1978SoPh...58..149T};
\opencite{1979AN....300..151S};
\opencite{1979AcASn..20..374M};
\opencite{1980AcASn..21..136Y};
\opencite{1980AcASn..21..152S}, \citeyear{1982SoPh...75..229S};
\opencite{1985SoPh...96..307S};
\opencite{1985AcApS...5...19L};
\opencite{1988SoPh..117...57Y};
\opencite{1990CoPhC..59..139L};
\opencite{1990AN....311..309L};
\opencite{1992ApJ...385..344K};
\opencite{1994SoPh..150..221D};
\opencite{1994ApJ...422..899M};
\opencite{1995A&A...298..277Y};
\opencite{1995A&A...303..927M};
\opencite{1997SoPh..174..311J};
\opencite{1999SoPh..188..345C};
\opencite{1999SoPh..190..107D};
\opencite{2000ApJ...540.1143Y};
\opencite{2000ApJ...543L..89W};
\opencite{yan01}, \citeyear{2001SoPh..201..337Y};
\opencite{2002ApJ...572..580W};
\opencite{2003SSRv..107..111Y};
\opencite{2004JKAS...37...41M};
\opencite{2004A&A...420..719F};
\opencite{2004ApJ...611..545G};
\opencite{2004A&A...423.1119B};
\opencite{2006SoPh..234...95D};
\opencite{2007A&A...475.1081L};
\opencite{2008ApJ...673L.207N};
\opencite{2008ApJ...676L..81J}, \citeyear{2009ApJ...696...84J},
\citeyear{2010ApJ...713..440J}, \citeyear{2012ApJ...752L...9J};
\opencite{2008ApJ...687..658W};
\opencite{2008SoPh..251..613M};
\opencite{2008A&A...488L..71T};
\opencite{2008ApJ...679.1629G};
\opencite{2009A&A...493..629Z};
\opencite{2009ApJ...704..341S}, \citeyear{2009ApJ...691..105S};
\opencite{2009ApJ...693..886D}
\opencite{2009SoPh..258...53C};
\opencite{2010A&A...510A..40H};
\opencite{2010ApJ...709L.142X};
\opencite{2010ApJ...716L..68C};
\opencite{2010ApJ...720.1102P};
\opencite{2010ApJ...721L.193L};
\opencite{2010ApJ...725L..38G};
\opencite{2011ApJ...732...87C};
\opencite{2011ApJ...738..161I};
\opencite{2012SoPh..276..133G};
\opencite{2012ApJ...750...12S};
\opencite{2012SoPh..278..367V};
\opencite{2012SoPh..278..411W};
\opencite{2012ApJ...759....1G};
\opencite{2012SoPh..281...53T}).

\section{Improvements and Challenges} \label{sec:disc}

	\subsection{Extrapolation Techniques}

Potential and LFF extrapolations are now mature techniques that are commonly used,
and their physical meaning and the understanding of physical process are well
established: no or limited amount of twist available in these models, and minimum
energy state, for instance. However, the NLFF model is still in constant
evolution/improvement. Since more than 20 years now, the NLFF assumption has been
implemented in the solar community to study a large number of topics with a
relatively good success. The next stages of development are:
\begin{itemize}
\item[-]{In Cartesian geometry, developing a reliable magnetohydrostatic
equilibrium code. This corresponds to a step forward compared to the force-free
field as far as the corona only is concerned. Including the thermodynamic
properties of the photosphere and/or chromosphere is a challenge.}
\item[-]{In spherical geometry, improving the spatial resolution of the PFSS
model is a first step, especially multigrids or nonuniform grids can be used.
The PFSS model also currently lacks information at the poles: the
improvement of the spatial resolution will need to be combined with a more
accurate measure of the radial magnetic field at the poles. This will modify the
dipole component of the magnetic field, which is needed to study the
dynamo action.}
\item[-]{In spherical geometry, despite the efforts reported in previous the
sections to move from the PFSS model to a NLFF assumption, the
use of the NLFF model could become a standard in the community with a special
care taken to the transition between the coronal magnetic field and the solar
wind magnetic field. As for the NLFF
field in Cartesian coordinates, a community effort should be envisioned.}
\item[-]{Defining contraints obtained from observations (X-rays, EUV, radio,
infrared, ...) to derive a magnetic field configuration closer to reality, and
thus to allow for quantitative comparison. As an example, the comparison between
the density derived from MHS models and the density obtained from spectrometers
should be a systematic check of the goodness of the extrapolation.}
\end{itemize}
The main word for the development of extrapolation techniques using solar data
as boundary conditions is quantitative. The main issue is always to find a
compromise between the spatial resolution, the size of the field-of-view, the
computational time, and the resources, which is the most suitable for the
physical problem tackled. 

	\subsection{Comments on Boundary Conditions} \label{sec:Cbound}
	
I briefly discuss the current issues encountered when using solar magnetograms
as boundary conditions for magnetic field extrapolations. In the following, a
``flat surface" refers to a surface of constant curvature or radius ({\em e.g.}, a
plane or a sphere). The main issues that should be kept in mind when
reconstructing a coronal magnetic field are:
\begin{itemize}
\item[-]{Inversion of Stokes parameters: the magnetic field components are
derived from the radiative transfer equations for the four Stokes parameters.
The inversion is complex and often used by reconstruction modellers as a black
box. One typical assumption is to consider that the solar atmosphere as a
constant optical depth. This assumption with the assumption of a local
thermodynamic equilibrium is becoming less and less accurate for high resolution
data.}
\item[-]{180 degree ambiguity: after the inversion of the Stokes parameters, an
ambiguity of 180 degrees on the orientation of the transverse magnetic field
still remains. Some algorithms to resolve this ambiguity are based on a magnetic
field model (potential, LFF, or NLFF field). The goodness of the ambiguity is
crucial, especially when a flux rope is present in the magnetic configuration.}
\item[-]{Flatness of the surface: as implied by the inversion of the Stokes
parameters, the measurement of the magnetic field in neighbouring pixels is
certainly not at the same height in the photosphere. This fact is even more
critical for chromospheric measurements. In order to currently perform a
magnetic field extrapolation, it is assumed that the magnetogram corresponds to
a flat surface.} 
\item[-]{The $\alpha$-distribution: the definition of $\alpha$ as seen in
Equation.~(\ref{eq:alpha}) is given by the normal component of the force-free equation.
It means that a contribution from the electric currents perpendicular to the
magnetic field is included into the $\alpha$-distribution when the magnetic
field is not strictly normal to the photospheric surface. The effects of the
electric currents have to be investigated in more details. For extrapolation
models beyond the force-free models, the components of the electric current
density will be more appropriate boundary conditions \cite{gra58}.}
\item[-]{Error estimate: it is still difficult to estimate the error/inacurracy
made in measuring the magnetic field: (i) the signal in sunspot is weak (limited
number of photons), (ii) the magnetic field in quiet Sun regions is weak, (iii)
the Hanle effect can dominate the quiet-Sun field, (iv) the linear polarisation
signal leading to the transverse magnetic field is small compared to the Stokes
$V$ signal (by an order of magnitude), (v) the properties of the spectral line
such as the Land\'e factor can influence the threshold/saturation of the
detected magnetic field strength, (vi) a pixel is not filled uniformly by
magnetic field, so the filling factor is also an important quantity. Error
estimates from vector magnetograms can be incorporate in the extrapolation model
resulting in a reliable equilibrium \cite{2011ApJ...728..112W}.}
\item[-]{Projection effect: the transverse field is strongly influenced by the
projection effect, especially in the penumbra of sunspots where the magnetic
field is mostly horizontal. This leads to the change of sign of observed
polarities.}
\item[-]{Synoptic/Carrington maps: often by default, the Carrington maps have
been corrected for the magnetic flux imbalance, which is good for extrapolation
but do not represent correctly the physics of the solar surface magnetic field.
}
\end{itemize}

In the near future, we will see the development of new instruments which will
provide us access to the magnetic field near the poles ({\em e.g.}, Solar Orbiter) and
in the corona ({\em e.g.}, CoMP). These improved measurements will impose new
constraints on the physics of magnetic field extrapolations.

\section{Summary} \label{sec:concl}

In this review, I have summarised many different results obtained from magnetic
field extrapolations combined with magnetic field observations. The
extrapolations have been applied to various different issues in order to
better understand the physics of the corona. Owing to the large number of
observations in strong field regions, most of the studies have dealt with the
structure and evolution of the magnetic field in active regions, as well as
studying the causes and consequences of flares and CMEs. 

What have we learned from magnetic field extrapolations? 
\begin{itemize}
\item[-]{The current force-free models describe relatively well the structure of
the coronal magnetic above active regions.}

\item[-]{The amount of magnetic energy available is consistent with what is
expected and observed during eruptive events, even if most of time, the studies
do not specify error bars.}

\item[-]{The existence of twisted flux bundles which are a prime ingredient in
most of the MHD models to store magnetic energy in the corona, and trigger
eruptions.}

\item[-]{The physics of flares is closely related to the topology of the
magnetic field prior to the eruption: existence of coronal null points,
separators or quasi-spearatrix layers.}

\item[-]{The large-scale connectivity of the magnetic field lines deduced from
extrapolations is crucial to understanding of the redistribution of magnetic energy
and magnetic helictiy in the solar corona.}

\item[-]{The link between the broad range of observations (from hard X-ray to
radio wavelengths) can be made through the structure of the magnetic field.}

\end{itemize}

What have we not learned (yet) from magnetic field extrapolations? 
\begin{itemize}
\item[-]{The ``universal" ingredient responsible for triggering
eruptions and which will allow us to predict flares and CMEs.}

\item[-]{The interaction between the different spatial scales involved in the
solar atmosphere, from the
quiet-Sun magnetic field to the global structure.}

\item[-]{The interaction between the different regions of the Sun: from the
tachocline to the solar wind, englobing the whole heliosphere.}

\item[-]{To be specific to a particular extrapolation method, we do not yet
understand the goodness of the force-free extrapolations while the imposed
boundary conditions are not force-free. To achieve the goal, it is required to
develop more theoretical studies based on the effects of noise or perpendicular
currents on force-free magnetic configurations.}
\end{itemize}

Despite this apparent success of extrapolation methods, {\em the improvement in
the understanding of the coronal magnetic field will depend on quantitative
comparisons with the observations} and not just qualitative comparisons. The
quantitative success of the extrapolation method will {\em a posteriori} justify
the physical assumptions or will drive the developements of these methods
towards more physical and sophisticated techniques. The limits of these
developments are the boundary conditions obtained by the observations, and the
computational power available. Especially, there is a need to improve the
physics incorporated into these models: the coupling between the
plasma and the magnetic field playing a major role in the evolution of the
magnetic fields.

%
\begin{acks}
I would like to thank Steve Tomczyk and Marc DeRosa for inviting me to
contribute to the "Coronal Magnetism -- Connecting Models to Data and the Corona
to the Earth" workshop held in Boulder (2012). I also thank Michael Thompson for
his hospitality at HAO (Boulder) during the workshop. 
\end{acks}

%
%
\bibliographystyle{spr-mp-sola-cnd} 
%
%
%
%

\end{article} 
\end{document}